\documentclass[floatfix,aps,prl,preprint, superscriptaddress]{revtex4-2}

\usepackage{upgreek}
\usepackage{graphicx}
\usepackage[]{epsfig}
\usepackage{times,amsmath,amssymb}
\usepackage{color}
\usepackage[colorlinks = true,
            linkcolor = blue,
            urlcolor  = blue,
            citecolor = blue,
            anchorcolor = blue]{hyperref}

\begin{document}

\title{Dynamics of quantum cellular automata electron transition in triple quantum dots}

\author{Takumi Aizawa}
\affiliation{Research Institute of Electrical Communication, Tohoku University, 2-1-1 Katahira, Aoba-ku, Sendai 980-8577, Japan}
\affiliation{Graduate School of Engineering, Tohoku University, 468-1 Aramaki Aza Aoba, Aoba-ku, Sendai 980-0845, Japan}

\author{Motoya Shinozaki}
\affiliation{WPI Advanced Institute for Materials Research, Tohoku University, 2-1-1 Katahira, Aoba-ku, Sendai 980-8577, Japan}

\author{Yoshihiro Fujiwara}
\affiliation{Research Institute of Electrical Communication, Tohoku University, 2-1-1 Katahira, Aoba-ku, Sendai 980-8577, Japan}
\affiliation{Graduate School of Engineering, Tohoku University, 468-1 Aramaki Aza Aoba, Aoba-ku, Sendai 980-0845, Japan}

\author{Takeshi Kumasaka}
\affiliation{Research Institute of Electrical Communication, Tohoku University, 2-1-1 Katahira, Aoba-ku, Sendai 980-8577, Japan}

\author{Wataru Izumida}
\affiliation{Department of Physics, Tohoku University, 6-3, Aramaki Aza-Aoba, Aoba-ku, Sendai 980-8578, Japan}

\author{Arne Ludwig}
\affiliation{Lehrstuhl f\"{u}r Angewandte Festk\"{o}rperphysik, Ruhr-Universit\"{a}t Bochum, D-44780 Bochum, Germany}

\author{Andreas D. Wieck}
\affiliation{Lehrstuhl f\"{u}r Angewandte Festk\"{o}rperphysik, Ruhr-Universit\"{a}t Bochum, D-44780 Bochum, Germany}

\author{Tomohiro Otsuka}
\email[]{tomohiro.otsuka@tohoku.ac.jp}
\affiliation{WPI Advanced Institute for Materials Research, Tohoku University, 2-1-1 Katahira, Aoba-ku, Sendai 980-8577, Japan}
\affiliation{Research Institute of Electrical Communication, Tohoku University, 2-1-1 Katahira, Aoba-ku, Sendai 980-8577, Japan}
\affiliation{Graduate School of Engineering, Tohoku University, 468-1 Aramaki Aza Aoba, Aoba-ku, Sendai 980-0845, Japan}
\affiliation{Center for Spintronics Research Network, Tohoku University, 2-1-1 Katahira, Aoba-ku, Sendai 980-8577, Japan}
\affiliation{Center for Science and Innovation in Spintronics, Tohoku University, 2-1-1 Katahira, Aoba-ku, Sendai 980-8577, Japan}
\affiliation{Center for Emergent Matter Science, RIKEN, 2-1 Hirosawa, Wako, Saitama 351-0198, Japan}

\date{\today}

\begin{abstract}
The quantum cellular automata (QCA) effect is a transition in which multiple electron move coordinately by Coulomb interactions and observed in multiple quantum dots.
This effect will be useful for realizing and improving quantum cellular automata and information transfer using multiple electron transfer.
In this paper, we investigate the real-time dynamics of the QCA charge transitions in a triple quantum dot by using fast charge-state readout realized by rf reflectometry.
We observe real-time charge transitions and analyze the tunneling rate comparing with the first-order tunneling processes.
We also measure the gate voltage dependence of the QCA transition and show that it can be controlled by the voltage.
\end{abstract}

\maketitle

Semiconductor quantum dots, as controllable artificial quantum systems~\cite{Tarucha1996prl, Ciorga2000prb, Raymond2004prl, kouwenhoven2001few}, are possible candidates of quantum bits for quantum information processing~\cite{Loss1998pra, Petta2005science, koppens2006driven, hanson2007spins, ladd2010quantum, yoneda2014fast, veldhorst2015two, takeda2016fault, otsuka2016single, yoneda2018quantum}.
Beyond these well-known applications, there is also growing anticipation for computational methods that utilize the quantum cellular automata (QCA) effect.
The QCA has attracted much attention due to its small feature size approaching the atomic scale, and its low power consumption, being one of the applications for conducting wiring-free cell network~\cite{tougaw1994logical, lent1993quantum, Islamshah1999science}, as well as that itself can be applied to quantum information processing~\cite{Toth2001pra}.

The QCA effect, characterized by the simultaneous transfer of several electrons, is realized through Coulomb interaction in a system of coupled quantum dots. 
Such a system can be demonstrated by using semiconductor quantum dots~\cite{otsuka2016single, ito2016detection, ito2018four}, and the QCA effect has also been observed in triple quantum dots~\cite{Gaudreau2006prl, Schroer2007prb}.
However, the QCA effect has been observed only in measurements on charge state diagrams representing static equilibrium states, and the details of its electron transport dynamics have not yet been detailed.
On the other hand, the electron tunneling between remote quantum dots has been investigated, and the contribution of virtual intermediate states is obtained~\cite{braakman2013long}.
This suggests the existence of a similar contribution in QCA charge transitions.

To probe the dynamics of the QCA effect, we employ a fast time-resolved technique known as radio-frequency (rf) reflectometry~\cite{Schoelkopf1998science, Reilly2007apl, Barthel2009prl, Barthel2010prb}.
We form triple quantum dots with charge states realizing the QCA effect and observe its dynamics.
Additionally, we monitor the dynamics of charge transitions, focusing on sequential electron transport that corresponds to the virtual intermediate states of QCA charge transitions.
We demonstrate that the dependence of the QCA effect on gate voltage can be explained by considering the influence of gate voltage on the virtual intermediate states.

\begin{figure}
  \centering
  \includegraphics[scale=1]{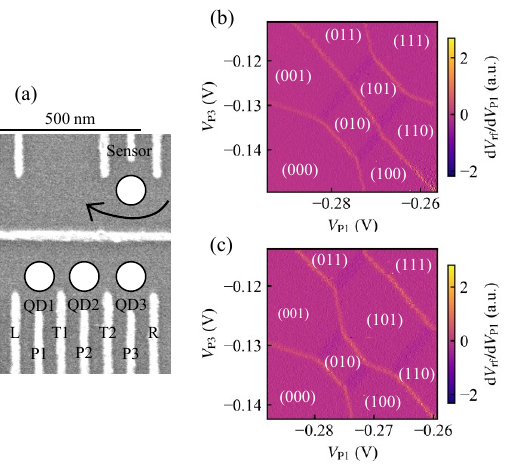}
  \caption{(a) Scanning electron micrograph of the device. The charge state of the triple quantum dot is monitored by the quantum dot sensor connected to the high-frequency resonator circuit.
  (b) Charge stability diagram around (010)-(101) transition including quadruple points showing the charge sensing signal $V_{\rm rf}$ as a function of $V_{\rm P1}$ and $V_{\rm P3}$. The number of electrons in each quantum dot is shown as ($n_{1}$ $n_{2}$ $n_{3}$). (c) Charge stability diagram around (010)-(101) transition including triple points.}
  \label{fig1}
\end{figure}

Figure~\ref{fig1}(a) shows a scanning electron micrograph of the device. 
The device is fabricated on a GaAs/AlGaAs heterostructure wafer by depositing Ti/Au gate electrodes on the surface. The white structures correspond to the gate electrodes. 
The upper dot corresponds to a quantum dot charge sensor and the bottom to the target triple quantum dot (QD1, QD2, and QD3), respectively.
The charge state of QD1, QD2 and QD3 is monitored by the quantum dot charge sensor. 
The sensor is connected to a high-frequency tank circuit for the rf reflectometry (the resonance frequency $f_{\rm res}=176$~MHz), and information on the charge state of the triple quantum dot is extracted from the reflected rf signal.
The detail of the measurement setup is described in Ref.~\cite{shinozaki2021gate}.
All measurements are conducted in a dilution refrigerator with a base temperature of 50 mK. 

Figure~\ref{fig1}(b) displays the charge stability diagram, illustrating the charge sensing signal $V_{\rm rf}$ as a function of $V_{\rm P1}$ and $V_{\rm P3}$. 
The number of electrons in each quantum dot is denoted as ($n_{1}$ $n_{2}$ $n_{3}$).
The QCA charge transition is expected at the boundary between (010) and (101) states.
To simplify the measurement, we modify the charge stability diagram by applying slightly negative $V_{\rm P2}$, transitioning from Fig.~\ref{fig1}(b) to (c).
As a result, the quadruple points involving the four states (101), (010), (001), (011), and (101), (010), (100), (110) transform into triple points comprising three states: (101), (010), (001), and (101), (010), (100) respectively.

\begin{figure}
  \centering
  \includegraphics{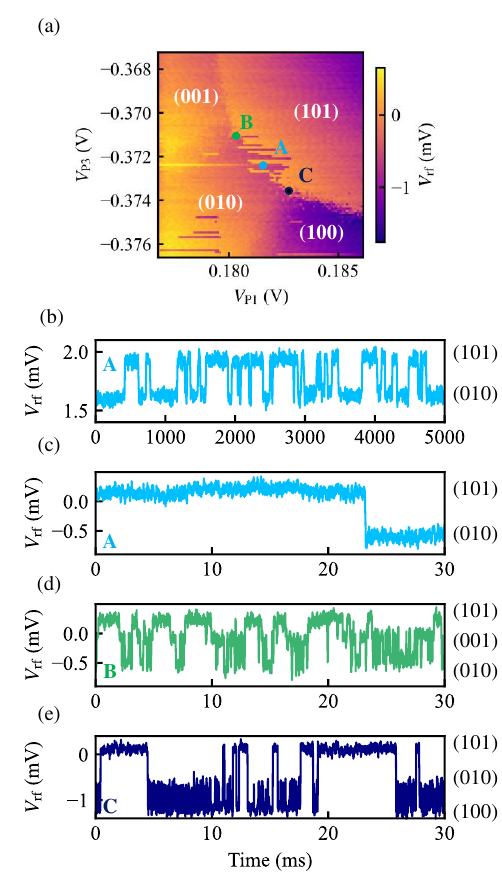}
  \caption{(a) An expanded stability diagram around the (010)-(101) charge transition line.  (b) Results of real-time charge sensing measurements for 5000~ms with a sampling rate of 1 MHz on condition A. 
  (c), (d), and (e) Faster time traces for 30~ms with the sampling rate of 125 MHz on conditions A, B, and C, respectively.}
  \label{fig2}
\end{figure}

Figure~\ref{fig2}(a) shows an expanded stability diagram around the (010)-(101) charge transition. 
To probe the dynamics of the electron tunneling for the real time measurement by the rf reflectometry, we adjust the tunneling barriers in the triple dot.
This tuning shifts the position of the charge transition lines from Fig.~\ref{fig1}(c) and induces the scattering of the charge transition lines in Fig.~\ref{fig2}(a) by the slow tunneling rates compared to the sweep of $V_{\rm P1}$.
Figures~\ref{fig2}(b)-(e) present the results of real-time charge sensing measurements corresponding to each of the voltage conditions A, B, and C depicted in Figure~\ref{fig2}(a).
For Figs.~\ref{fig2}(c), (d), and (e), the sampling rate is set at $125$ MHz, while for Fig.~\ref{fig2}(b), it is $1$ MHz. 
The raw data measured at these rates are subsequently averaged over $1000$ and $5000$ points, respectively, for each sampling rate.
Figures~\ref{fig2}(b) shows that a slow binary fluctuation is observed between (010) and (101) states in condition A, where the QCA charge transitions occur. 
Even in the case of the faster measurement, as shown in Fig.~\ref{fig2}(b), no intermediate state (either (001) or (100)) is observed, and it can be seen that two electrons tunnel simultaneously in the QCA charge transition.
Note that the difference in $V_{\rm rf}$ between Figs.~\ref{fig2}(b) and (c) results from changes in the charge sensor condition.
On the other hand, Figures~\ref{fig2}(d) and (e) show that in conditions B and C near the triple point, which are out of condition A where QCA charge transitions occur, triple transitions are observed.
Specifically, these transitions occur between (010)-(001)-(101) in condition B and (100)-(010)-(101) in condition C, respectively.
It indicates that single-electron tunneling occurs in these regions, and this tunneling process is faster than the QCA charge transition.
In these conditions, the occurrence of the QCA charge transition is rare, indicating that it is not the dominant process.
Note that the change in the tunneling barrier between quantum dots due to the small difference of the gate voltages $V_{\rm P1}$ and $V_{\rm P3}$ for conditions A, B, and C is considered negligible.

\begin{figure}
  \centering
  \includegraphics{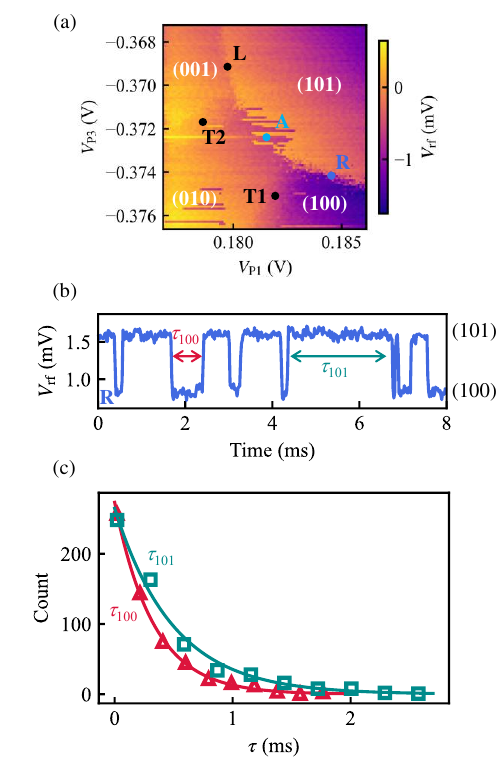}
  \caption{(a) An expanded stability diagram around the (010)-(101) charge transition line with new measurement points. (b) A result of real-time charge sensing measurements for 8~ms on condition R. (c) Observed distributions of $\tau_{100}$ and $\tau_{101}$ on condition R.}
  \label{fig3}
\end{figure}

Next, we analyze the detail of the tunneling rates.
The QCA charge transition is characterized as a second-order tunneling process, involving the simultaneous tunneling of two electrons. 
This process is likely associated with the related first-order tunneling process.
The first-order single electron tunneling occurs on conditions L, T1, T2, and R.
We conduct measurements of the real-time charge transition on such conditions L, T1, T2, and R in Fig.~\ref{fig3}(a).
Figure~\ref{fig3}(b) shows the real-time charge transition trace on condition R. 
A binary fluctuation is observed between states of (100)-(101). 
Let $\tau_{100}$ and $\tau_{101}$ be the retention time in each condition. 
The $\tau_{100}$ and $\tau_{101}$ are counted by repeating the measurement and their distributions are shown in Fig.~\ref{fig3}(c). 
The distribution follows an exponential decay $\exp (-\varGamma_{i}\tau_{i})$ where $i=(100)$ or $(101)$~\cite{schleser2004time, gustavsson2009electron, Guttinger2011prb, Otsuka2017srep, Otsuka2019prb}.
The fitting results are indicated by red and cyan lines in Fig.~\ref{fig3}(c). 
$\varGamma_{(100)}$ and $\varGamma_{(101)}$ are obtained as fitting parameters. 
For the quantitative analysis of the tunneling rate at condition R $\varGamma_{R}$, the relation in the following equation is used~\cite{schleser2004time, gustavsson2009electron, braakman2013long}.
\begin{equation}
 \varGamma_{R} = \varGamma_{(100)}+\varGamma_{(101)}
 \label{eq1}
\end{equation}
By substituting $\varGamma_{(100)}$ and $\varGamma_{(101)}$ obtained from the fitting results into Eq.~\ref{eq1}, $\varGamma_{R}=5.1 \times 10^3$ Hz is obtained. 
The results of evaluating the tunneling rates by performing the same analysis for each of the conditions A, L, T1, T2, and R are summarized in Table~\ref{table1}.
The QCA tunneling rate $\varGamma$ observed at the condition A is smaller than the first-order tunneling process $\varGamma_{\mathrm{L}}$, $\varGamma_{\mathrm{T1}}$, $\varGamma_{\mathrm{T2}}$, and $\varGamma_{\mathrm{R}}$.

\begin{table}
  \caption{The evaluated QCA tunneling rate and the first-order tunneling rates}
  \label{table1}
  \vspace{2mm}
  \centering
  \begin{tabular}{c|c}
  Tunneling process & Tunneling rate (Hz) \\
  \hline\hline
  $\varGamma$ & $1.4 \times 10^1$  \\
  \hline
  $\varGamma_{\mathrm{L}}$ & $9.9 \times 10^3$ \\
  \hline
  $\varGamma_{\mathrm{T1}}$ & $3.8 \times 10^4$ \\
  \hline   
  $\varGamma_{\mathrm{T2}}$ & $2.4 \times 10^4$ \\
  \hline   
  $\varGamma_{\mathrm{R}}$ & $5.1 \times 10^3$ \\
  \end{tabular}
\end{table}

From Fermi's golden rule, the relation between the tunnel rate of the QCA charge transition $\varGamma$  and the tunnel rate of the first-order processes $\varGamma_{\mathrm{L}}$, $\varGamma_{\mathrm{T1}}$, $\varGamma_{\mathrm{T2}}$, $\varGamma_{\mathrm{R}}$ becomes the follwoing.
\begin{equation}
  \varGamma = \frac{\hbar \gamma}{2} \Bigg\{ \frac{\varGamma_{\mathrm{L}}\varGamma_{\mathrm{T2}}}{(\varepsilon_{3}-\varepsilon_{2})^2}+\frac{\varGamma_{\mathrm{T1}}\varGamma_{\mathrm{R}}}{(\varepsilon_{1}-\varepsilon_{2})^2} 
  +\frac{\varGamma_{\mathrm{L}}\varGamma_{\mathrm{T2}}}{(U_{12}-U_{13}+\varepsilon_{2}-\varepsilon_{3})^2}+\frac{\varGamma_{\mathrm{T1}}\varGamma_{\mathrm{R}}}{(U_{23}-U_{13}+\varepsilon_{2}-\varepsilon_{1})^2} \Bigg\}
\label{eq2}
\end{equation}
where $\hbar$ is Planck's constant, $\gamma $ is the broadening of the dot level, $\varepsilon_{i}$ is the dot level energy and $U_{ij}$ is the Coulomb energy between quantum dots. 
There are four possible virtual intermediate states in the QCA charge transition, (100), (001), (110), and (011), corresponding to the first, second, third, and fourth terms in Eq.~\ref{eq2}, respectively. 
Here, we assume that the change of the tunneling rate by the change of the plunger's operation point is negligible because the change of the plunger gate voltages are small.
The energy in the denominator of Eq.~\ref{eq2} should vary with gate voltage. 
Therefore, the QCA tunneling rate $\varGamma$ should have a dependence on the gate voltage.

\begin{figure}
  \centering
  \includegraphics{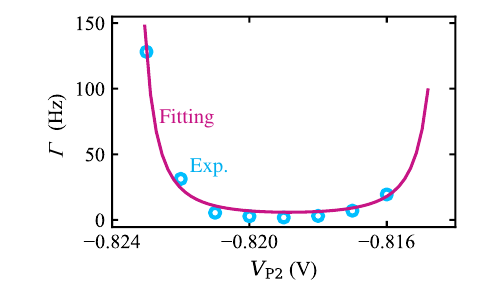}
  \caption{Evaluated tunnel rate of the QCA charge transition $\varGamma$ as a function of the gate voltage $V_{\mathrm{P2}}$ (circles). The pink trace shows the result of the fitting.}
  \label{fig4}
\end{figure}

Figure~\ref{fig4} shows the measured QCA tunneling rate $\varGamma$ as a function of the gate voltage $V_{\mathrm{P2}}$. 
Clearly seen, $\varGamma$ has a dependence on the  $V_{\mathrm{P2}}$. 
The Eq.~\ref{eq2} can be rewritten by using $V_{\mathrm{P2}}$ as the following equation.
\begin{equation}
  \varGamma = \frac{\hbar \gamma }{2} (\varGamma_{\mathrm{L}}\varGamma_{\mathrm{T2}}+\varGamma_{\mathrm{T1}}\varGamma_{\mathrm{R}}) 
  \Bigg\{ \frac{1}{b^2(V_{\mathrm{P2}}-a)^2}+\frac{1}{b^2(c-(V_{\mathrm{P2}}-a))^2} \Bigg\}
\label{eq3}
\end{equation}
where $a$ is a parameter indicating the energy offset, $b$ is the conversion factor between voltage and energy, and $c$ is the voltage representation of the Coulomb energy between quantum dots, respectively.
Here, we assume $\varepsilon_{2}-\varepsilon_{3} = \varepsilon_{2}-\varepsilon_{1}$ and $U_{12}=U_{23}$.
The values of $\varGamma_{\mathrm{L}}\varGamma_{\mathrm{T2}}+\varGamma_{\mathrm{T1}}\varGamma_{\mathrm{R}}$ are directly substituted for the results in Table~\ref{table1}. 
The result of the fitting with Eq.~\ref{eq3}, using fitting parameters of $a$, $b$, and $c$, is shown as a pink trace in Fig.~\ref{fig4}.
Equation~\ref{eq3} reproduces the experimental results well, supporting that the observed transitions are due to the second order QCA effect, and the tunneling rate can be controlled by the gate voltage.

In conclusion, we reveal the real-time dynamics of the QCA charge transitions, which are multi-electron transfers due to Coulomb interactions, by using fast charge-state readout realized by rf reflectometry. 
We conduct an analysis of the tunneling rate, comparing the QCA charge transitions with first-order tunneling processes. 
Furthermore, the presented gate voltage dependence of the QCA transition rate reveals results that align well with the theoretical model, demonstrating that it can be modulated by the gate voltage. 
Our findings hold promise for the realization of quantum cellular automata and information transfer through multiple electron transfers.

The authors thank T. Nakajima, J. Yoneda, K. Takeda, A. Noiri, S. Tarucha,
and RIEC Fundamental Technology Center and the Laboratory for Nanoelectronics and Spintronics
for fruitful discussions and technical support. 
Part of this work is supported by 
MEXT Leading Initiative for Excellent Young Researchers, 
Grants-in-Aid for Scientific Research (21K18592, 23H01789, 23H04490), 
JFE 21st Century Foundation Research Grant, 
Tanigawa Foundation Research Grant, 
The Foundation for Technology Promotion of Electronic Circuit Board,
Iketani Science and Technology Foundation Research Grant,
The Ebara Hatakeyama Memorial Foundation Research Grant,
and FRiD Tohoku University.

\bibliography{reference.bib}

\end{document}